\documentclass{sigchi}

\toappear{Permission to make digital or hard copies of all or part of this work for personal or classroom use is      granted without fee provided that copies are not made or distributed for profit or commercial advantage and that copies bear this notice and the full citation on the first page. Copyrights for components of this work owned by others than ACM must be honored. Abstracting with credit is permitted. To copy otherwise, or republish, to post on servers or to redistribute to lists, requires prior specific permission and/or a fee. Request permissions from permissions@acm.org. \\
{\emph{CHI'14}}, April 26--May 1, 2014, Toronto, Canada. \\
Copyright \copyright~2014 ACM ISBN/14/04...\$15.00. \\}
\usepackage{balance}  % to better equalize the last page
\usepackage{graphics} % for EPS, load graphicx instead 
\usepackage[T1]{fontenc}
\usepackage{txfonts}
\usepackage{mathptmx}
\usepackage[pdftex]{hyperref}
\usepackage{color}
\usepackage{booktabs}
\usepackage{textcomp}
\usepackage{microtype} % Improved Tracking and Kerning
\usepackage{ccicons}  % Cite your images correctly!
\usepackage{todonotes}

\def\plaintitle{ The effect of self-efficacy and pair programming experience in learning results of introductory programming courses}

\def\emptyauthor{}
\def\plainkeywords{Human-Computer Interaction; Self-efficacy; Pair Programming; Collaborative Learning}

\makeatletter
\def\url@leostyle{%
  \@ifundefined{selectfont}{
    \def\UrlFont{\sf}
  }{
    \def\UrlFont{\small\bf\ttfamily}
  }}
\makeatother
\def\pprw{8.5in}
\def\pprh{11in}

\definecolor{linkColor}{RGB}{6,125,233}
\hypersetup{%
  pdftitle={\plaintitle},
  pdfauthor={\emptyauthor},
  pdfkeywords={\plainkeywords},
  bookmarksnumbered,
  pdfstartview={FitH},
  colorlinks,
  citecolor=black,
  filecolor=black,
  linkcolor=black,
  urlcolor=linkColor,
  breaklinks=true,
}

\begin{document}

\title{\plaintitle}

\numberofauthors{3}  
\author{%
  \alignauthor{Yifan Mei\\
    \affaddr{Department of Statistics, UW-Madison}\\
    \affaddr{Madison, USA}\\
    \email{ymei8@wisc.edu}}\\
  \alignauthor{Heng Ping\\
    \affaddr{iSchool,\\UW-Madison}\\
    \affaddr{Madison, USA}\\
    \email{hping2@wisc.edu}}\\
  \alignauthor{Mingren Shen\\
    \affaddr{Biophysics, \\UW-Madison}\\
    \affaddr{Madison, USA}\\
    \email{mshen32@wisc.edu}}\\
}

\maketitle

\begin{abstract}
  %UPDATED---\today.

The purpose of this study was to explore the interactive effect of self-efficacy and pair programming experience to the final learning results in introductory programming courses. We developed a 2x2 fractional design to explore their roles and relationships. Data was collected by distributing questionnaires to students have learnt or are learning CS367 at UW-Madison. They were asked to evaluate their self-efficacy levels and pair programming experience. After that, they needed to complete a quiz of 11 Java knowledge quiz indicating their learning results. We present results from 36 participants which show that students with high self-efficacy levels tended to earn a higher score in the Java knowledge quiz.  However, pair programming experience shows no significant effects on learning results.Our finding suggests that high self-efficacy levels have a positive impact on students' performance in introductory programming courses.

\end{abstract}

\category{H.5.m.}{[Information Interfaces and Presentation
  (e.g. HCI)]}{Miscellaneous} 
  \category{K.3.2.}{[Computer and Education]}{Computer and Information Science Education}

\keywords{\plainkeywords}

\section{Introduction}
Computer education is important for the development modern society. According to data from White house \footnote{\url{https://obamawhitehouse.archives.gov/blog/2016/01/30/computer-science-all}}, there were more than 600,000 high-paying computer science-related jobs across the United States that were unfilled in 2016. Computer science and data science are not only important for the technology companies, but also for so many industries, including transportation, health-care, education, and financial services. Although both global governments and companies like Facebook, Google and Microsoft have been spending large amount of energy ,money and human resources for computer science education throughout the past few years, the dropout and failure rates in introductory programming courses are still very high. Some studies even suggest that the dropout and failure rates could be as high as 30 percent \cite{guzdial2002teaching}.  However, for some,  learning programming is both enjoyable and motivating while others find it a miserable struggle they fail to complete.

There have been many studies of the factors that may influence beginners' success in introductory programming course \cite{bergin2005programming,wilson2001contributing}. These factors include previous computing experience \cite{hagan2000does}, computer self-efficacy \cite{karsten1998computer}, mathematics or science background \cite{wilson2001contributing}, teaching method like pair-programing \cite{mcdowell2002effects,de2016pair} and student personal attributes like learning style and mental models \cite{ramalingam2004self,wilson2001contributing}.

The goal of this project is to investigate the effect of self-efficacy and pair programming experiences to the study results of introductory level programmers, explore the relationship between these two very important concepts and study their interactions on students' learning performance.

\section{Related Work}

According to Bandura, self-efficacy is defined as 

\begin{quote}
People's beliefs about their capabilities to produce designated levels of performance that exercise influence over events that affect their lives \cite{bandura1997self}. 
\end{quote}

Researchers have recognized self-efficacy as an essential enhancement of learning motivations \cite{zimmerman2000self}. Scales of measuring self-efficacy are developed to assess people's self-efficacy levels \cite{sherer1982self}. In a two-year research conducted in North Carolina State University, a pair programming experiment was executed to evaluate students' self-efficacy in an introductory computer science course \cite{williams2001support}. The result shows that positive self-efficacy perception would encourage students to perform more actively in programming practice \cite{kinnunen2011cs}. 

Previous research has shown that working in pairs would enable students to improve their programming skills as well as final scores than individually programming learning \cite{williams2000all,mcdowell2002effects,mcdowell2003impact,sun2020assessing,awe2022machine,gurbani2021evaluation,liu2019harmonization,ming2016mesoscale,field2020rapid,shen2019diffusive,morganautomated}, Therefore, as a proven efficient pedagogy methodology, pair programming is widely utilized in introductory computer courses in the Uinted States. Yet there are still issues that might cause pair programming to fail. For example, in a Microsoft paper, cost-efficiency and personality conflicts are reported as two top questions disturbing pair programming \cite{begel2008pair}.

Self-efficacy is recognized as a significant internal factor that influencing new skill learning \cite{schumacher2013developing} and pair programing is believed a good learning method for introductory courses \cite{mcdowell2002effects,de2016pair,luo2018n, shen2021deep,shen2021multi,jacobs2022performance,shen2016chemically,field2021development,shen2023machine,lawrence2020exploring,shen2022query,shen2021machine}. However, the interaction between self-efficacy and pair programming has not been assessed, and this is the question we interested most in our project.

\section{Methodology}
Our design is a two way between - participants design. We collect our data by distributing questionnaires, each of which has 3 parts to measure self-efficacy levels, pair programming experience and learning results of CS introductory programming courses respectively. The participants are students who have learned or are learning CS367 at UW-Madison. We use 2 x 2 factorial design to study the influence of self-efficacy and pair-programming experience on the learning results of CS introductory programming courses. The two independent variables, self-efficacy level and pair programming experience are both categorical variables with 2 levels, low (bad) or high (good). 

The model for the fractional design is shown as following:

\begin{equation}
\label{Fractional}
Y_{ijl} = \mu + \alpha_i + \beta_j + (\alpha \beta)_{ij} + \varepsilon_{ijl}
\end{equation}

In equation~\eqref{Fractional}, Y represents the dependent variable, the learning results of CS introductory programming course. In our project, this dependent variable is measured by scores from a Java knowledge quiz. $\alpha$ is the first independent variable, self-efficacy levels;  $\beta$ is the second independent variable, pair programming experience; ($\alpha \times \beta$) represents the interaction term of self-efficacy levels and pair programming experience. The meaning of subscripts in equation~\eqref{Fractional} is given below:

\begin{itemize}
\item i = 1, 2; representing the 2 levels of self-efficacy, low and high; 
\item j = 1, 2; representing the 2 levels of pair programming experience, bad and good;
\item l = 1, 2,$\dots$, 9; representing the 9 individual participants in each 2 x 2 block
\item $\varepsilon_{ijl} \sim  N(0, \sigma_{\varepsilon_{ijl}})$; representing the error term of the model
\end{itemize}

And the hypotheses for our designs are as following:

\begin{enumerate}
\item H1: The higher the self-efficacy levels, the better the learning results of CS introductory programming courses. 
\item H2: The better the pair programming experience, the better the learning results of CS introductory programming courses. 
\item H3: High self-efficacy but good pair programming still leads to good learning results.
\item H4: Low self-efficacy but bad pair programming leads to bad learning results.
\end{enumerate}

\section{Measurement}
The measurement in our study included a self-efficacy self-evaluation, a pair programming experience survey and a Java knowledge quiz.

Self-efficacy levels was measured using the items and key points from International Personality Item Pool(IPIP)\footnote{\url{http://ipip.ori.org/newindexofscalelabels.htm}} and questions from
previous studies \cite{ramalingam1998development,wilson2001contributing}. The scale of self-efficacy contains 10 questions helping participants judge their self-efficacy in a wide range of programing tasks and situations, e.g. "I felt confident in my ability during the programming tasks.", "I could come up with good solutions when programming." and "I do not see the consequence of my program",etc. Responses are on a 7-points Likert scale, where 1 means strongly disagree, and 7 means strongly agree.

Participants' pair programming experience was evaluated by using questions used by Salleh et al. \cite{salleh2014investigating}. The purpose of their study was to investigate the effects of personality traits on pair programming and we modifying their questions to suit our purpose to test the participants' pair programming satisfaction scale based on their personal experience. Sample questions are like "My motivation level increased when working with my partner(Working with a partner makes it easy to understand what I am doing and why I am doing it)." ,"My level of confidence in solving the exercises increased when working with my partner." and "Do you switch roles working in pair programming(change "driver" and "navigator" fairly regularly)?", etc. We also used 7-points Likert scale survey items ranging form 1 (bad) to 7 (good).

For measuring the learning results of CS introductory programming course, we prepared 11 multiple questions from Java Review website\footnote{\url{http://interactivepython.org/runestone/static/JavaReview/index.html}}. Java Review website is a wonderful, free and open website providing questions for Java learners to test their Java knowledge. It is created by Prof. Barbara Ericson at Georgia Tech.  The Java knowledge quiz contains questions from different topics like "Java Basics","Classes and Objects" and "Strings", etc. Within each topic, the questions are separated by 3 different levels: basic, intermediate, and hard. In our design, around 70\% of the questions are chosen from the basic level, around 20\% of the questions are chosen from the intermediate level, and the rest question is hard level.

\subsection{Procedure}
This study was conducted during the course of CS 770 in UW-Madison. And we distributed the questionnaire that contains 3 parts of the above 32 questions by Google form\footnote{\url{https://docs.google.com/forms/d/e/1FAIpQLSf4tpQ5G2PpgMqj0he0EpliJaHNlK5zVPaoBMCx2fbIy6Q9xA/viewform?usp=sf_link}}. All participants are required to finish all the 3 parts of the survey which normally took 20 to 30 minutes.

We collected 36 valid questionnaires. Then, we summed up each students' the points for each part respectively. For the self-efficacy level and pair programming experience, we transformed the raw points in the questions to get the self-efficacy or pair programming experience scale points. For negative workload items, we converted the value using the max value(7 in our scales) - raw point. 

\begin{figure}
\centering
  \includegraphics[width=0.6\columnwidth]{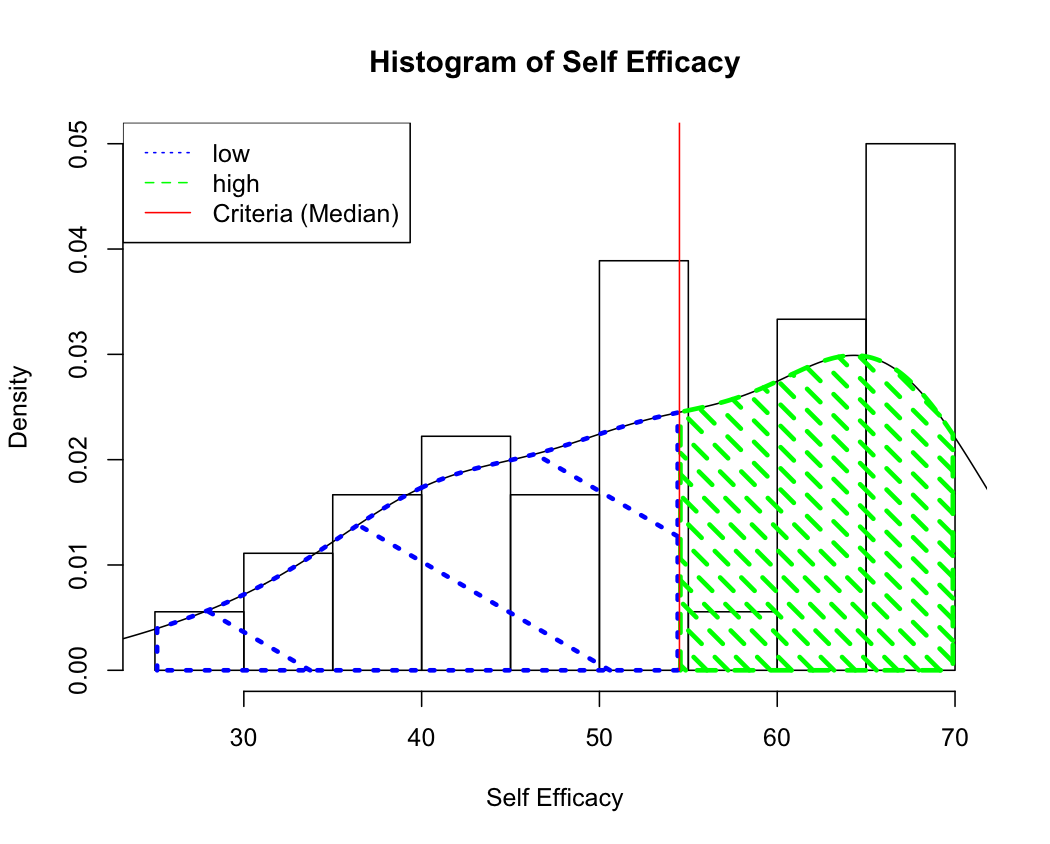}
  \caption{Distribution of self-efficacy points and changing self-efficacy points into two factor levels low and high }~\label{fig:figure1}
\end{figure}

Then, based on the summarized self-efficacy scale points, we first transformed the sum of self-efficacy points into two levels, low and high. To do so, we sorted the sum points of self-efficacy for the 36 participants by ascending order and treated the top 18 participants as low level self-efficacy participants, while the rest 18 participants as high-level self-efficacy participants as shown in Fig.\ref{fig:figure1}.

\begin{figure}
\centering
  \includegraphics[width=0.6\columnwidth]{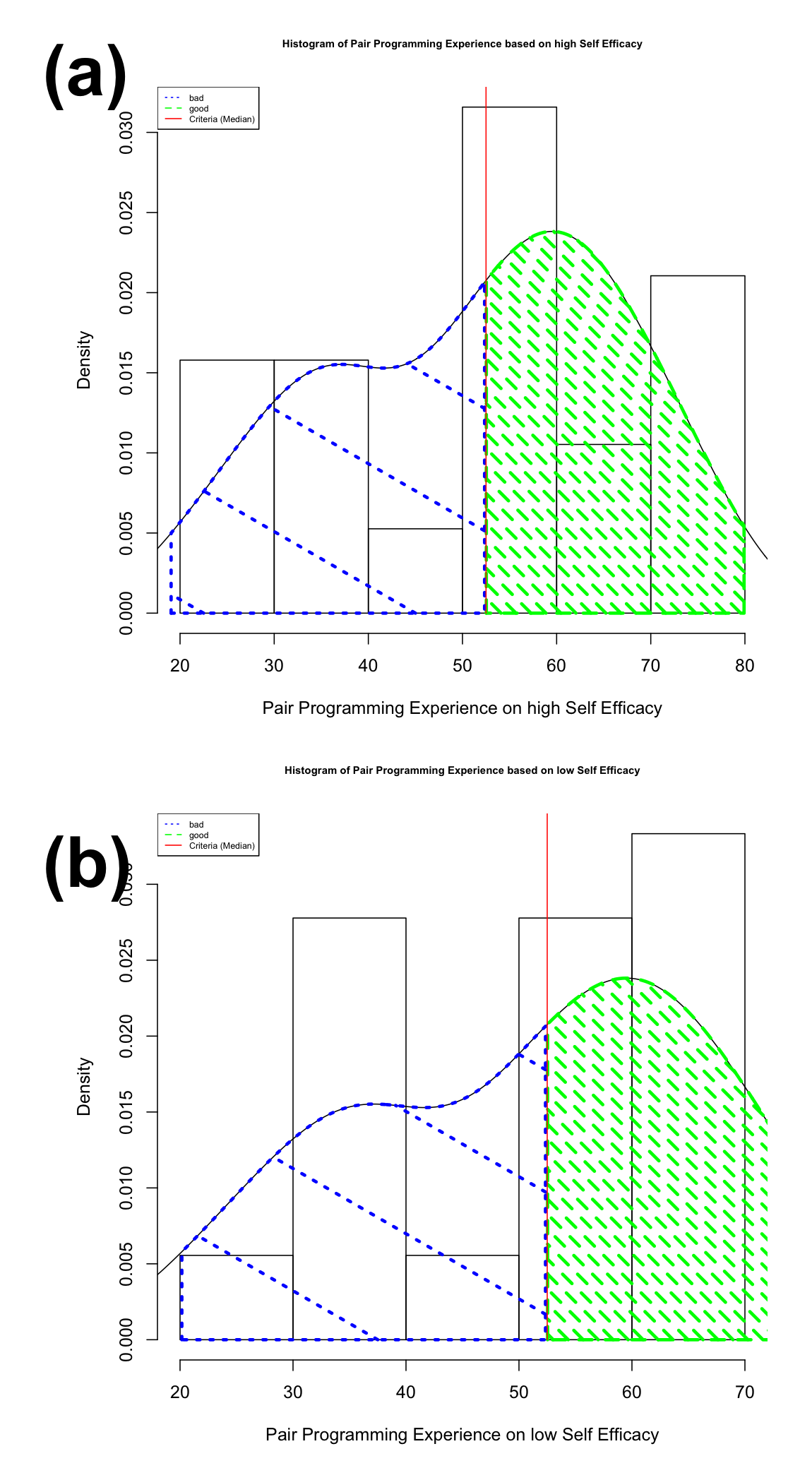}
  \caption{(a)Distribution of pair programming experience points in high self-efficacy group and changing pair programming experience points  into two factor levels bad and good.(b)Distribution of pair programming experience points in low self-efficacy group and changing pair programming experience points  into  factor levels bad and good }~\label{fig:figure2}
\end{figure}

Then, we sorted the 2 groups of low and high self-efficacy( 18 participants each ) in increasing order separately by the sum of pair programming experience scores, and treated the top 9 of the participants in each group had bad pair programming experience, while the rest 9 of each group had good pair programming experience as shown in Fig.\ref{fig:figure2}. 

\section{Results}

After data collection, we got a 2x2 factorial design data set, where the first independent variable is self-efficacy with 2 levels: low and high, while the second independent variable is pair programming experience with 2 levels: bad, good. Within each 2 x 2 block, there were 9 participants. What we could do next is to make 2 bar-plots for the 2 treatments, and get some general information from the bar-plots.

\begin{figure}
\centering
  \includegraphics[width=0.6\columnwidth]{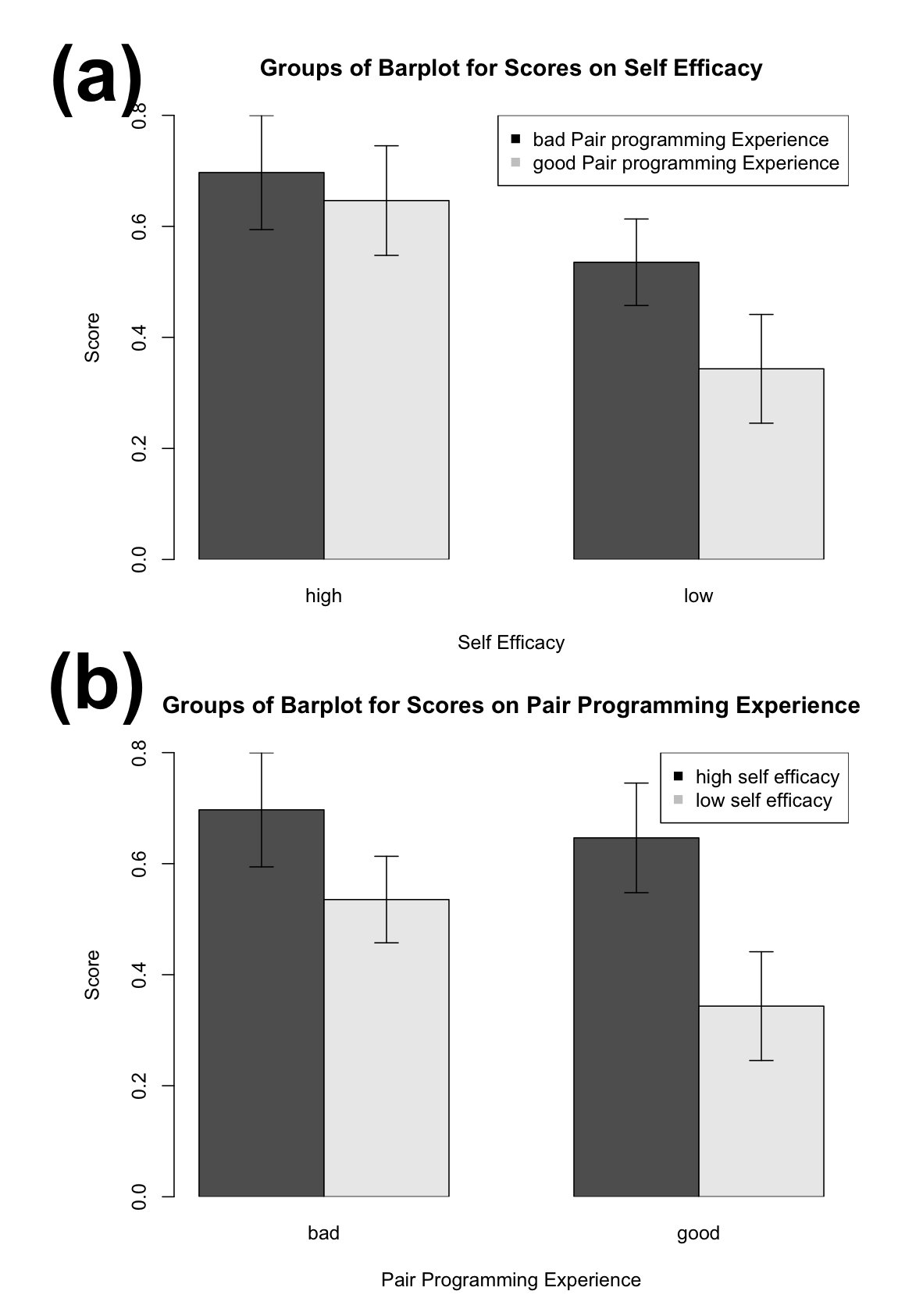}
  \caption{(a)Barplots for scores on self-efficacy groups. (b)Barplots for scores on pair programing experience groups.}~\label{fig:figure3}
\end{figure}

According to the Fig.\ref{fig:figure3}, we can see that the each subgroup has a similar standard error. Besides, people who have relatively bad pair programming experience get higher scores no matter their self-efficacy was high or low, while high self-efficacy resulted in high scores. 

\begin{figure}
\centering
  \includegraphics[width=0.6\columnwidth]{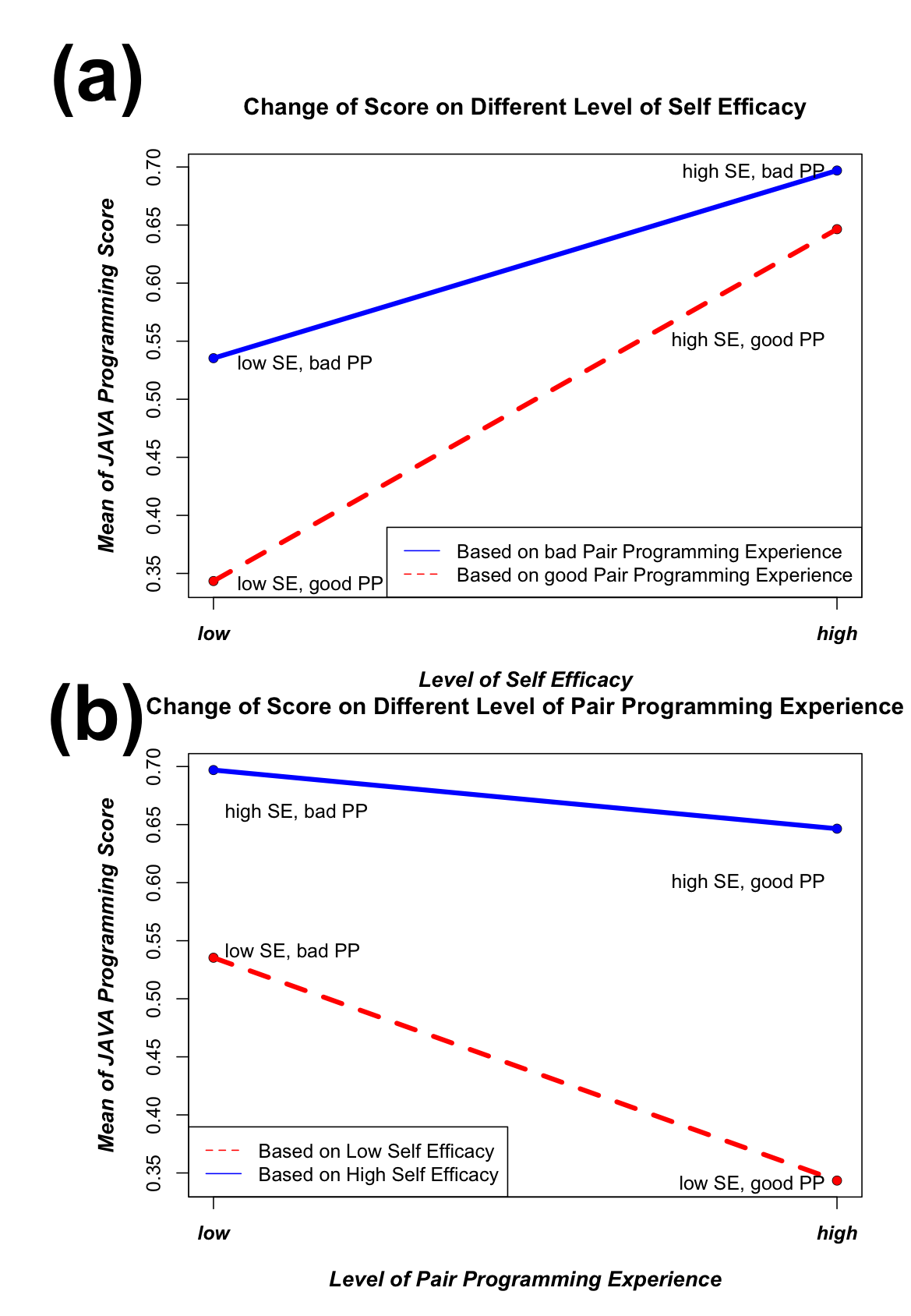}
  \caption{(a)Influence of different self-efficacy levels. (b) Influence of different levels of pair programing.}~\label{fig:figure4}
\end{figure}

After we got some sense about the main effects of the model, our next step was to make a general plot to see whether there were any potential interactions between the 2 independent variables, self efficacy levels and pair programming experience. The results are shown in Fig.~\ref{fig:figure4}.

According to the interaction detecting plots in Fig.\ref{fig:figure4}, there was no obvious interaction existed in any of the graph. Besides, the interaction plots showed the same result as bar plot, where no matter self-efficacy was high or low, people with worse pair programming experience have higher scores, and no matter pair programming experience was good or bad, higher self-efficacy leaded to better scores.

Now, after some general views of the main effects and interaction effects, we calculated the ANOVA table to get a more precise test about the interactions and relationships between the 2 independent variables. The result was shown in Table.\ref{table1} .

\begin{table}[ht]
\begin{center}
\caption{ \textit{Two Way Between-Participants ANOVA Table of self-efficacy and pair programming experience}}
\label{table1}
\begin{tabular}{lrrrrc}
  \hline
\textit{ Factors} & \textit{df} & \textit{SS}& \textit{MS} & \textit{F} & \textit{p} \\ 
  \hline
SE      &1	&0.49	&0.49	 &6.00	  &0.020$^*$  \\
PPE         &1	&0.13	&0.13	 &1.63	  &0.21      \\ 
 SE \& PPE          &1	&0.045	&0.045	&0.56	&0.46  \\
  Residuals  	&32	&2.59 & 0.081 &  &  \\ 
   \hline
   \hline
\multicolumn{6}{l}{Signif. codes    0 \textbf{ *** } 0.001 \textbf{ ** } 0.010\textbf{ * } 0.050 \textbf{ . } 0.10 \textbf{   } 1} \\ 
\hline \\
\multicolumn{6}{l}{\small SE : self-efficacy; PPE :  pair programming experience} \\
\multicolumn{6}{l}{\small SE \& PPE  : interactions between self-efficacy and pair programming experience.}
\end{tabular}
\end{center}
\end{table}

Based on the ANOVA table(Table.\ref{table1}), we could find that we only had weak evidences to say that the self-efficacy levels have positive influences on Java knowledge quiz scores meaning that we had weak evidences to say that the higher the level of self-efficacy, the better the learning result of CS introductory programming courses. As for the experience of pair programming or the interactions between pair programming experience and self-efficacy levels, we did not have any evidence to say they had influences on the learning results of CS introductory programming courses.

The result here partly fitted our hypothesis. For the four hypotheses we proposed, only H1 was supported by our data which meant that the higher the self-efficacy levels, the better the learning results of CS introductory programming courses. 

\section{Discussion}
\subsection{self-efficacy}
According to previous research results \cite{ramalingam2004self}, self-efficacy has been recognized as a motivation factor in college-level learning. High self-efficacy level leads to strong desire in competition with other students and inspires students to earn higher scores in the Java programming quiz which agrees with our project results. 

So in general, we can propose that self-efficacy is more important for people to perform well in introductory programming courses.

\subsection{pair programming experience}
It seems that higher level of pair programming experience did not contribute to a high score in the Java programming quiz. Many factors might lead to this results. Firstly, to full utilize the power of pair programming, all learners should receive proper training, such as switching roles frequently and the steering keyboard and mouse alternatively. Such a complex human-human interactions like pair programming requires both parties put great effort to communicate, discuss and cooperate with each other.

Secondly, some participants with advanced programming background usually do not rely on pair programming or cooperate well with their pair programming partner. In another word, the individual learning process is efficient and effective enough for them. This could be attributed to sampling bias when distributing questionnaires.

Finally, pair programming is aimed at improving students’ coding ability instead of their examination scores \cite{de2016pair}. However, as we used a multiple choice Java knowledge quiz to assess the learning results, there could be a possibility that some participants are not good at doing multiple choice quiz but good at coding.

\section{Conclusion}
Our study focused on testing the main effects and interaction effects of self-efficacy level and pair programming experience to the learning result of CS introductory programming courses. We conducted a 2 x 2 factorial design for the design, and we found out that self-efficacy was the only significant factor that influencing the final learning results, whereas pair programming experience and the interaction effects of self-efficacy and pair programming experience had no significant influence on the final learning results.

\section{Acknowledgments}

We thank all the volunteers, all classmates and professors who supported and provided helpful suggestions on our project. 

We especially want to thank Prof. Mutlu at UW-Madison for instructing our whole project and helpful suggestions and discussions,  Prof. Legault at UW-Madison for helping us prepare the questionnaire used in our study and Prof. Deppeler at UW-Madison for helping us recruit the volunteers in CS 367.We also would like to thank Prof. Barbara Ericson at Georgia Tech for the Java questions we used from her wonderful Java Review Website\footnote{\url{http://interactivepython.org/runestone/static/JavaReview/index.html}}.

% REFERENCES FORMAT
% References must be the same font size as other body text.
\bibliographystyle{SIGCHI-Reference-Format}
\bibliography{sample}

%%% -*-BibTeX-*-
%%% Do NOT edit. File created by BibTeX with style
%%% ACM-Reference-Format-Journals [18-Jan-2012].

\begin{thebibliography}{00}

%%% ====================================================================
%%% NOTE TO THE USER: you can override these defaults by providing
%%% customized versions of any of these macros before the \bibliography
%%% command.  Each of them MUST provide its own final punctuation,
%%% except for \shownote{}, \showDOI{}, and \showURL{}.  The latter two
%%% do not use final punctuation, in order to avoid confusing it with
%%% the Web address.
%%%
%%% To suppress output of a particular field, define its macro to expand
%%% to an empty string, or better, \unskip, like this:
%%%
%%% \newcommand{\showDOI}[1]{\unskip}   % LaTeX syntax
%%%
%%% \def \showDOI #1{\unskip}           % plain TeX syntax
%%%
%%% ====================================================================

\ifx \showCODEN    \undefined \def \showCODEN     #1{\unskip}     \fi
\ifx \showDOI      \undefined \def \showDOI       #1{{\tt DOI:}\penalty0{#1}\ }
  \fi
\ifx \showISBNx    \undefined \def \showISBNx     #1{\unskip}     \fi
\ifx \showISBNxiii \undefined \def \showISBNxiii  #1{\unskip}     \fi
\ifx \showISSN     \undefined \def \showISSN      #1{\unskip}     \fi
\ifx \showLCCN     \undefined \def \showLCCN      #1{\unskip}     \fi
\ifx \shownote     \undefined \def \shownote      #1{#1}          \fi
\ifx \showarticletitle \undefined \def \showarticletitle #1{#1}   \fi
\ifx \showURL      \undefined \def \showURL       #1{#1}          \fi

\bibitem{awe2022machine}
{Adam~M Awe}, {Michael~M Vanden~Heuvel}, {Tianyuan Yuan}, {Victoria~R Rendell},
  {Mingren Shen}, {Agrima Kampani}, {Shanchao Liang}, {Dane~D Morgan}, {Emily~R
  Winslow}, {and} {Meghan~G Lubner}. 2022.
\newblock \showarticletitle{Machine learning principles applied to CT radiomics
  to predict mucinous pancreatic cysts}.
\newblock {\em Abdominal Radiology\/} (2022), 1--11.
\newblock


\bibitem{bandura1997self}
{Albert Bandura}. 1997.
\newblock {\em Self-efficacy: The exercise of control}.
\newblock Macmillan.
\newblock


\bibitem{begel2008pair}
{Andrew Begel} {and} {Nachiappan Nagappan}. 2008.
\newblock \showarticletitle{Pair programming: what's in it for me?}. In {\em
  Proceedings of the Second ACM-IEEE international symposium on Empirical
  software engineering and measurement}. ACM, 120--128.
\newblock


\bibitem{bergin2005programming}
{Susan Bergin} {and} {Ronan Reilly}. 2005.
\newblock \showarticletitle{Programming: factors that influence success}. In
  {\em ACM SIGCSE Bulletin}, Vol.~37. ACM, 411--415.
\newblock


\bibitem{de2016pair}
{Carolina~Alves de Lima~Salge} {and} {Nicholas Berente}. 2016.
\newblock \showarticletitle{Pair Programming vs. Solo Programming: What Do We
  Know After 15 Years of Research?}. In {\em System Sciences (HICSS), 2016 49th
  Hawaii International Conference on}. IEEE, 5398--5406.
\newblock


\bibitem{field2021development}
{Kevin~G Field}, {Ryan Jacobs}, {Mingen Shen}, {Matthew Lynch}, {Priyam Patki},
  {Christopher Field}, {and} {Dane Morgan}. 2021.
\newblock \showarticletitle{Development and deployment of automated machine
  learning detection in electron microcopy experiments}.
\newblock {\em Microscopy and Microanalysis\/} {27}, S1 (2021), 2136--2137.
\newblock


\bibitem{field2020rapid}
{Kevin~G Field}, {Mingren Shen}, {Caleb~P Massey}, {Kenneth~C Littrell}, {and}
  {Dane~D Morgan}. 2020.
\newblock \showarticletitle{Rapid Characterization Methods for Accelerated
  Innovation for Nuclear Fuel Cladding}.
\newblock {\em Microscopy and Microanalysis\/} {26}, S2 (2020), 868--869.
\newblock


\bibitem{gurbani2021evaluation}
{Sidharth Gurbani}, {Dane Morgan}, {Varun Jog}, {Leo Dreyfuss}, {Mingren Shen},
  {Arighno Das}, {E~Jason Abel}, {and} {Meghan~G Lubner}. 2021.
\newblock \showarticletitle{Evaluation of radiomics and machine learning in
  identification of aggressive tumor features in renal cell carcinoma (RCC)}.
\newblock {\em Abdominal Radiology\/}  {46} (2021), 4278--4288.
\newblock


\bibitem{guzdial2002teaching}
{Mark Guzdial} {and} {Elliot Soloway}. 2002.
\newblock \showarticletitle{Teaching the Nintendo generation to program}.
\newblock {\it Commun. ACM} {45}, 4 (2002), 17--21.
\newblock


\bibitem{hagan2000does}
{Dianne Hagan} {and} {Selby Markham}. 2000.
\newblock \showarticletitle{Does it help to have some programming experience
  before beginning a computing degree program?}. In {\em ACM SIGCSE Bulletin},
  Vol.~32. ACM, 25--28.
\newblock


\bibitem{jacobs2022performance}
{Ryan Jacobs}, {Mingren Shen}, {Yuhan Liu}, {Wei Hao}, {Xiaoshan Li}, {Ruoyu
  He}, {Jacob~RC Greaves}, {Donglin Wang}, {Zeming Xie}, {Zitong Huang}, {and}
  {others}. 2022.
\newblock \showarticletitle{Performance and limitations of deep learning
  semantic segmentation of multiple defects in transmission electron
  micrographs}.
\newblock {\em Cell Reports Physical Science\/} {3}, 5 (2022).
\newblock


\bibitem{karsten1998computer}
{Rex Karsten} {and} {Roberta~M Roth}. 1998.
\newblock \showarticletitle{Computer self-efficacy: A practical indicator of
  student computer competency in introductory IS courses}.
\newblock {\em Informing Science\/} {1}, 3 (1998), 61--68.
\newblock


\bibitem{kinnunen2011cs}
{P{\"a}ivi Kinnunen} {and} {Beth Simon}. 2011.
\newblock \showarticletitle{CS majors' self-efficacy perceptions in CS1:
  results in light of social cognitive theory}. In {\em Proceedings of the
  seventh international workshop on Computing education research}. ACM, 19--26.
\newblock


\bibitem{lawrence2020exploring}
{Nick Lawrence}, {Mingren Shen}, {Ruiqi Yin}, {Cloris Feng}, {and} {Dane
  Morgan}. 2020.
\newblock \showarticletitle{Exploring Generative Adversarial Networks for
  Image-to-Image Translation in STEM Simulation}.
\newblock {\em arXiv preprint arXiv:2010.15315\/} (2020).
\newblock


\bibitem{liu2019harmonization}
{Yilin Liu}, {Gregory~R Kirk}, {Brendon~M Nacewicz}, {Martin~A Styner},
  {Mingren Shen}, {Dong Nie}, {Nagesh Adluru}, {Benjamin Yeske}, {Peter~A
  Ferrazzano}, {and} {Andrew~L Alexander}. 2019.
\newblock \showarticletitle{Harmonization and targeted feature dropout for
  generalized segmentation: Application to multi-site traumatic brain injury
  images}. In {\em Domain Adaptation and Representation Transfer and Medical
  Image Learning with Less Labels and Imperfect Data: First MICCAI Workshop,
  DART 2019, and First International Workshop, MIL3ID 2019, Shenzhen, Held in
  Conjunction with MICCAI 2019, Shenzhen, China, October 13 and 17, 2019,
  Proceedings 1}. Springer, 81--89.
\newblock


\bibitem{luo2018n}
{Guan-Zheng Luo}, {Ziyang Hao}, {Liangzhi Luo}, {Mingren Shen}, {Daniela
  Sparvoli}, {Yuqing Zheng}, {Zijie Zhang}, {Xiaocheng Weng}, {Kai Chen},
  {Qiang Cui}, {and} {others}. 2018.
\newblock \showarticletitle{N 6-methyldeoxyadenosine directs nucleosome
  positioning in Tetrahymena DNA}.
\newblock {\em Genome biology\/}  {19} (2018), 1--12.
\newblock


\bibitem{mcdowell2002effects}
{Charlie McDowell}, {Linda Werner}, {Heather Bullock}, {and} {Julian Fernald}.
  2002.
\newblock \showarticletitle{The effects of pair-programming on performance in
  an introductory programming course}.
\newblock {\em ACM SIGCSE Bulletin\/} {34}, 1 (2002), 38--42.
\newblock


\bibitem{mcdowell2003impact}
{Charlie McDowell}, {Linda Werner}, {Heather~E Bullock}, {and} {Julian
  Fernald}. 2003.
\newblock \showarticletitle{The impact of pair programming on student
  performance, perception and persistence}. In {\em Proceedings of the 25th
  international conference on Software engineering}. IEEE Computer Society,
  602--607.
\newblock


\bibitem{ming2016mesoscale}
{Shen Ming-Ren}, {Liu Rui}, {Hou Mei-Ying}, {Yang Ming-Cheng}, {and} {Chen Ke}.
  2016.
\newblock \showarticletitle{Mesoscale simulation of self-diffusiophoretic
  microrotor}.
\newblock {\em ACTA PHYSICA SINICA\/} {65}, 17 (2016).
\newblock


\bibitem{morganautomated}
{Dane Morgan}, {Ryan Jacobs}, {Mingren Shen}, {Priyam Patki}, {Matthew Lynch},
  {and} {Kevin Field}.
\newblock \showarticletitle{Automated Defect Detection in Electron Microscopy
  of Radiation Damage in Metals}.
\newblock  (????).
\newblock


\bibitem{ramalingam2004self}
{Vennila Ramalingam}, {Deborah LaBelle}, {and} {Susan Wiedenbeck}. 2004.
\newblock \showarticletitle{Self-efficacy and mental models in learning to
  program}. In {\em ACM SIGCSE Bulletin}, Vol.~36. ACM, 171--175.
\newblock


\bibitem{ramalingam1998development}
{Vennila Ramalingam} {and} {Susan Wiedenbeck}. 1998.
\newblock \showarticletitle{Development and validation of scores on a computer
  programming self-efficacy scale and group analyses of novice programmer
  self-efficacy}.
\newblock {\em Journal of Educational Computing Research\/} {19}, 4 (1998),
  367--381.
\newblock


\bibitem{salleh2014investigating}
{Norsaremah Salleh}, {Emilia Mendes}, {and} {John Grundy}. 2014.
\newblock \showarticletitle{Investigating the effects of personality traits on
  pair programming in a higher education setting through a family of
  experiments}.
\newblock {\em Empirical Software Engineering\/} {19}, 3 (2014), 714--752.
\newblock


\bibitem{schumacher2013developing}
{Daniel~J Schumacher}, {Robert Englander}, {and} {Carol Carraccio}. 2013.
\newblock \showarticletitle{Developing the master learner: applying learning
  theory to the learner, the teacher, and the learning environment}.
\newblock {\em Academic Medicine\/} {88}, 11 (2013), 1635--1645.
\newblock


\bibitem{shen2021machine}
{Mingren Shen}. 2021.
\newblock {\em Machine Learning Applications in Material Science Problems}.
\newblock The University of Wisconsin-Madison.
\newblock


\bibitem{shen2022query}
{Mingren Shen}, {Shuoxuan Dong}, {and} {Xiuyuan He}. 2022.
\newblock \showarticletitle{Query Time Optimized Deep Learning Based Video
  Inference System}.
\newblock {\em arXiv preprint arXiv:2212.06905\/} (2022).
\newblock


\bibitem{shen2021multi}
{Mingren Shen}, {Guanzhao Li}, {Dongxia Wu}, {Yuhan Liu}, {Jacob~RC Greaves},
  {Wei Hao}, {Nathaniel~J Krakauer}, {Leah Krudy}, {Jacob Perez}, {Varun
  Sreenivasan}, {and} {others}. 2021a.
\newblock \showarticletitle{Multi defect detection and analysis of electron
  microscopy images with deep learning}.
\newblock {\em Computational Materials Science\/}  {199} (2021), 110576.
\newblock


\bibitem{shen2021deep}
{Mingren Shen}, {Guanzhao Li}, {Dongxia Wu}, {Yudai Yaguchi}, {Jack~C Haley},
  {Kevin~G Field}, {and} {Dane Morgan}. 2021b.
\newblock \showarticletitle{A deep learning based automatic defect analysis
  framework for In-situ TEM ion irradiations}.
\newblock {\em Computational Materials Science\/}  {197} (2021), 110560.
\newblock


\bibitem{shen2019diffusive}
{Mingren Shen}, {Rui Liu}, {Ke Chen}, {and} {Mingcheng Yang}. 2019.
\newblock \showarticletitle{Diffusive-flux-driven microturbines by fore-and-aft
  asymmetric phoresis}.
\newblock {\em Physical Review Applied\/} {12}, 3 (2019), 034051.
\newblock


\bibitem{shen2023machine}
{Mingren Shen}, {Dina Sheyfer}, {Troy~David Loeffler}, {Subramanian~KRS
  Sankaranarayanan}, {G~Brian Stephenson}, {Maria~KY Chan}, {and} {Dane
  Morgan}. 2023.
\newblock \showarticletitle{Machine learning for interpreting coherent X-ray
  speckle patterns}.
\newblock {\em Computational Materials Science\/}  {230} (2023), 112500.
\newblock


\bibitem{shen2016chemically}
{Mingren Shen}, {Fangfu Ye}, {Rui Liu}, {Ke Chen}, {Mingcheng Yang}, {and}
  {Marisol Ripoll}. 2016.
\newblock \showarticletitle{Chemically driven fluid transport in long
  microchannels}.
\newblock {\em The journal of chemical physics\/} {145}, 12 (2016).
\newblock


\bibitem{sherer1982self}
{Mark Sherer}, {James~E Maddux}, {Blaise Mercandante}, {Steven Prentice-Dunn},
  {Beth Jacobs}, {and} {Ronald~W Rogers}. 1982.
\newblock \showarticletitle{The self-efficacy scale: Construction and
  validation}.
\newblock {\em Psychological reports\/} {51}, 2 (1982), 663--671.
\newblock


\bibitem{sun2020assessing}
{Xiaoyu Sun}, {Nathaniel~J Krakauer}, {Alexander Politowicz}, {Wei-Ting Chen},
  {Qiying Li}, {Zuoyi Li}, {Xianjia Shao}, {Alfred Sunaryo}, {Mingren Shen},
  {James Wang}, {and} {others}. 2020.
\newblock \showarticletitle{Assessing graph-based deep learning models for
  predicting flash point}.
\newblock {\em Molecular Informatics\/} {39}, 6 (2020), 1900101.
\newblock


\bibitem{williams2001support}
{Laurie Williams} {and} {Richard~L Upchurch}. 2001.
\newblock \showarticletitle{In support of student pair-programming}. In {\em
  ACM SIGCSE Bulletin}, Vol.~33. ACM, 327--331.
\newblock


\bibitem{williams2000all}
{Laurie~A Williams} {and} {Robert~R Kessler}. 2000.
\newblock \showarticletitle{All I really need to know about pair programming I
  learned in kindergarten}.
\newblock {\it Commun. ACM} {43}, 5 (2000), 108--114.
\newblock


\bibitem{wilson2001contributing}
{Brenda~Cantwell Wilson} {and} {Sharon Shrock}. 2001.
\newblock \showarticletitle{Contributing to success in an introductory computer
  science course: a study of twelve factors}. In {\em ACM SIGCSE Bulletin},
  Vol.~33. ACM, 184--188.
\newblock


\bibitem{zimmerman2000self}
{Barry~J Zimmerman}. 2000.
\newblock \showarticletitle{Self-efficacy: An essential motive to learn}.
\newblock {\em Contemporary educational psychology\/} {25}, 1 (2000), 82--91.
\newblock


\end{thebibliography}

\end{document}